\documentclass[epj]{webofc}
\usepackage[utf8]{inputenc}
\usepackage[varg]{txfonts}   
\usepackage{booktabs}
\usepackage{xcolor}
\definecolor{darkred}{rgb}{0.4,0.0,0.0}
\definecolor{darkgreen}{rgb}{0.0,0.4,0.0}
\definecolor{darkblue}{rgb}{0.0,0.0,0.4}
\usepackage[bookmarks,linktocpage,colorlinks,
    linkcolor = darkred,
    urlcolor  = darkblue,
    citecolor = darkgreen]{hyperref}
\newcommand{\be}{\begin{equation}}
\newcommand{\ee}{\end{equation}}
\newcommand{\bea}{\begin{eqnarray}}
\newcommand{\eea}{\end{eqnarray}}
\newcommand{\bes}{\begin{eqnarray}}
\newcommand{\ees}{\end{eqnarray}}
\newcommand{\ba}{\begin{array}}
\newcommand{\ea}{\end{array}}

\newcommand{\Figc}[1]{Figure~}
\newcommand{\Figsc}[1]{Figures~}

\newcommand{\Eq}{Eq.~}

\newcommand{\Eqs}{Eqs.~}

\newcommand{\Fig}{Figure~}

\newcommand{\Tab}{Table~}

\newcommand{\Ref}{Ref.~}

\newcommand{\rmO}{\mathrm{O}}

\newcommand{\SU}[1]{$\mathrm{SU}(#1)$}

\newcommand{\Nc}{N}

\newcommand{\chidof}{\chi^2/\mathrm{dof}}

\newcommand{\Tr}{\text{Tr}}

\newcommand{\fm}{\mathrm{fm}}

\newcommand{\ev}[1]{\left\langle #1 \right \rangle}

\newcommand{\tHooft}{'t~Hooft }

\usepackage{subfigure}
\wocname{EPJ Web of Conferences}
\woctitle{Lattice2017}
\begin{document}
%
\selectlanguage{english}
\title{%
Large $N$ scaling and factorization
in SU($N$) Yang-Mills theory
}
\author{%
	\firstname{Miguel} \lastname{Garc\'ia Vera}\inst{1,2}\fnsep
	\thanks{Speaker, \email{miguel.garcia@desy.de}} \and
	\firstname{Rainer} \lastname{Sommer}\inst{1,2}
}
\institute{%
John von Neumann Institute for Computing (NIC), DESY, Platanenallee 6, D-15738 Zeuthen, Germany
\and
Institut für Physik, Humboldt Universität zu Berlin, Newtonstr. 15, D-12489 Berlin, Germany
}
\abstract{
We present results for Wilson loops smoothed with
the Yang-Mills gradient flow and matched through the scale $t_0$. They provide
renormalized and precise operators allowing to
 test the $1/N^2$ scaling both at finite lattice spacing
and in the continuum limit. Our results show an excellent scaling up to $1/N = 1/3$.
Additionally, we 
obtain a very precise non-perturbative confirmation of factorization in the large $N$ limit.
}
\hfill DESY 17-153

\maketitle
\section{Introduction}\label{intro}

A well known problem when studying QCD, is 
the fact that the coupling $g$ is in general not small at relevant energy 
scales. In \Ref\cite{'tHooft:1973jz}, \tHooft proposed to consider 
the \SU{\Nc} Yang-Mills gauge theory and use the inverse of the rank of the 
gauge group $1/\Nc$ as an expansion parameter instead of $g^2$. 
If the theory was solvable in the $1/\Nc \to 0$ limit, results at the physical 
value of $1/\Nc = 1/3$ could be recovered as corrections in powers of $1/\Nc$, and 
in the case of the pure gauge theory which we consider in this work, the 
expansion is in powers of $1/\Nc^2$. 

The fact that corrections to the large $\Nc$ limit are organized in a power 
series in $1/\Nc^2$ can be obtained in perturbation theory using the 
topological expansion proposed by `t~Hooft. In the full non-perturbative theory, 
however, this is  not a proven statement. In the past, several 
lattice collaborations have found agreement with  
scaling at a non-perturbative level~\cite{Teper:2008yi,Lucini:2012gg}. 
The precision of these studies was mainly limited for two reasons.
First, spectral quantities such as torelon and glueball masses 
were considered, which need extrapolations to large separations.  
Second the problem of a bulk phase transition at
intermediate lattice spacing and topological freezing at small 
lattice spacing are more and more severe at larger $N$ and make it difficult to reach the continuum limit\cite{DelDebbio:2002xa}. 
We overcome these difficulties by using 1) 
high precision results for smooth Gradient flow observables 
and 2) open boundary conditions~\cite{Luscher:2011kk}, extending the analysis of \cite{Ce:2016uwy} to several observables. 
In this way we are able to check large 
$\Nc$ scaling with excellent precision. 

One particular property of large $\Nc$ scaling is factorization. For any two local gauge invariant or Wilson loop operators $\mathcal{A}$ and $\mathcal{B}$, it is expected 
that~\cite{Makeenko:1979pb,Yaffe:1981vf}
\begin{equation}
	\ev{\mathcal{A}\mathcal{B}} = \ev{\mathcal{A}} \ev{\mathcal{B}} + \rmO(1/\Nc^2) \, ,
	\label{eq:factorization}
\end{equation}
which shows that correlation functions in the large $\Nc$ limit are given only 
as a product of the disconnected parts.  
\Eq\eqref{eq:factorization} is the basis for conceptual developments  
such as the master field 
\cite{Coleman:1980nk,Witten:1979pi}, and the idea of volume reduction
\cite{Eguchi:1982nm,GonzalezArroyo:2010ss}. In particular, the 
latter has been of considerable 
interest in the lattice community, as it implies that theories at larger
$\Nc$ can be simulated in smaller boxes, reducing the computational effort. 
We will verify factorization non-perturbatively and
with high precision.

\section{Observables}\label{obs}

As mentioned previously, our main observables are smooth Wilson loop operators. 
They are defined in terms of the gauge links smoothed with 
the Yang-Mills gradient flow~\cite{Narayanan:2006rf,Luscher:2010iy}. Using the flow 
one has access to renormalizable operators with a well defined and finite 
continuum limit~\cite{Luscher:2011bx,Lohmayer:2012ue}. 
Let us stress the fact that {\em without} the use of the gradient flow, 
the Wilson loops are affected by 
perimeter and corner divergences~\cite{Dotsenko:1979wb,Brandt:1981kf}. 
One needs to remove those, e.g. by taking Creutz ratios \cite{Creutz:1980hb}
before taking a continuum limit. 

\begin{figure}
   \centering
   \sidecaption
   \includegraphics[width=0.30\textwidth,clip]{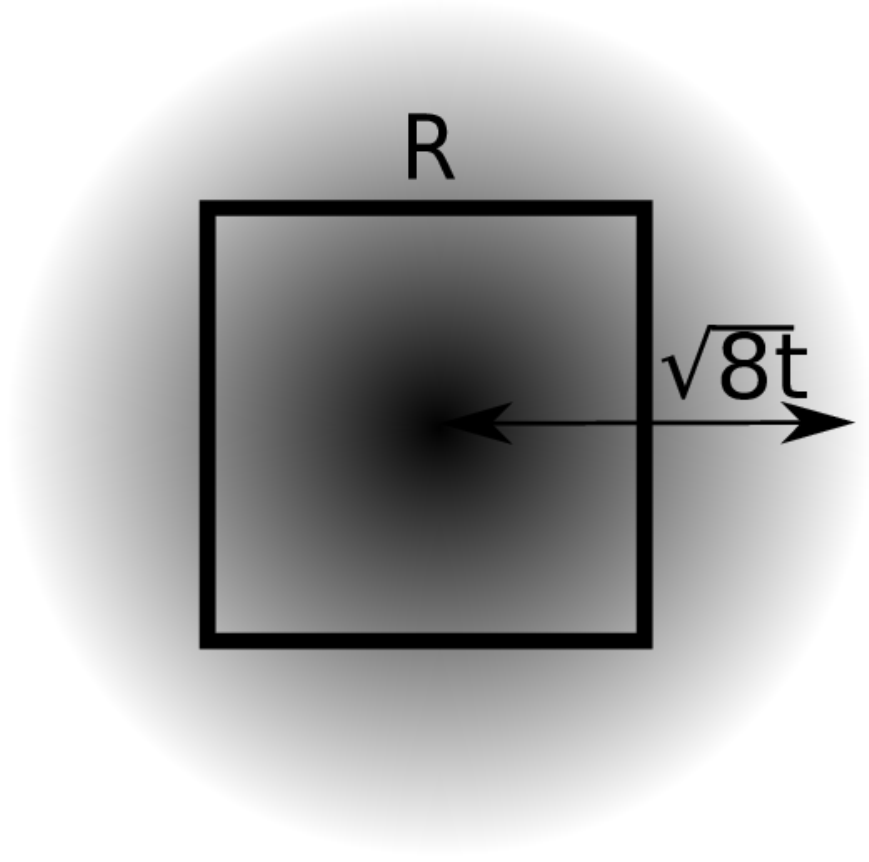}
   \caption{Graphical representation of a smooth Wilson loop. The side of the 
   loop $R$ is chosen so that it matches the smoothing radius of the flow, i.e. 
   $R = \sqrt{8t}$.}
   \label{fig:smoothloop}
\end{figure}

For a lattice of dimension $T \times L^3 $, with open boundary 
conditions in the time direction~\cite{Luscher:2011kk} and a 
lattice spacing $a$, we define the following operators 
\begin{align}\label{eq:looplattice}
W(c) &= 
 \frac{a^4}{\left( T - 2d \right) L^3} \sum_{x_0=d}^{T-d-a} 
 \sum_{\vec{x}} W(ct_0,x_0,\vec{x},R_c) \, ,\\\label{eq:loop2lattice}
W^2(c) &=  
 \frac{a^4}{\left( T - 2d \right) L^3} \sum_{x_0=d}^{T-d-a} 
 \sum_{\vec{x}} [W(ct_0,x_0,\vec{x},R_c)]^2 \, , \\\label{eq:Glattice}
 G_W(c) &= \frac{\ev{W^2(c)} - \ev{W(c)}^2}{\ev{W(c)}^2} \, ,
\end{align}
where $W(ct_0,x_0,\vec{x},R_c)$ is a square Wilson loop\footnote{Here we define a 
Wilson loop with the prefactor $1/\Nc$ such that it has a finite large $\Nc$ limit, 
i.e. $W(\mathcal{C}) = \frac{1}{\Nc} \Tr \, U(\mathcal{C})$, where  
$U(\mathcal{C})$ is the product of gauge links over the path $\mathcal{C}$.} in
a spatial plane with one corner 
at the point $(x_0,\vec{x})$ and a side of length $R_c = \sqrt{8ct_0}$. 
The parameter $d$ is chosen so that the effects of the open boundaries in the time 
direction are negligible with respect to the statistical error in the bulk~\cite{Ce:2016uwy}. 
Finally, $ct_0$ corresponds to the smoothing parameter, i.e. 
$t=ct_0$, with the scale $t_0$ defined for \SU{\Nc} as in \Ref\cite{Ce:2016awn}

\begin{equation}
	t^2 \ev{E(t)}|_{t=t_0} = 0.1125 \, \frac{\Nc^2 -1 }{\Nc} \, ,
	\label{eq:t0def}
\end{equation}
where $E(t)$ is the Yang-Mills action density computed through the clover definition 
on the lattice~\cite{Luscher:2010iy}. A schematic representation of a smooth Wilson loop is shown in 
\Fig\ref{fig:smoothloop}.

All the observables are measured for the gauge groups with $\Nc = 3,\, 4,\, 5,\, 6$ and 
$8$ on a set of lattices with a lattice spacing varying from $a \approx 0.1 \, \fm$ 
down to $a \approx 0.05 \, \fm$, and with a spatial size of 
$L \approx 1.55 \, \fm$.\footnote{We define physical units by $\sqrt{t_0}=0.166\,\fm$ 
motivated by its value in $\Nc=3$ derived from $r_0=0.5\fm$ 
\cite{Ce:2015qha,Sommer:1993ce}.
.}
We have simulated three or four lattice spacings for each gauge group, except 
for \SU{8} where we have a single lattice at $a \approx 0.08 \, \fm$. 
Details of the simulations will be presented in a forthcoming publication. 
The parameter $c$ in \Eqs\eqref{eq:looplattice}-\eqref{eq:Glattice} takes
three different values, $c=1/2, \, 1, \, 9/4$.

\section{Systematic effects}

As mentioned earlier, all our measurements are performed in lattices of approximately 
the same physical size $L \approx 1.55 \, \fm$. However, in order to make sure that 
the small mismatch (around $5\%$ in the worst case) between the different ensembles 
does not introduce a significant finite volume correction, we 
simulated two extra lattices, one for \SU{4} and one for \SU{5} at $L \approx 2.35 \, \fm$. 
These results are in perfect agreement with those at the smaller lattices for all the 
observables introduced in the previous section, including the Yang-Mills action density.

\begin{figure}
	\centering
	\includegraphics[width=0.49\textwidth]{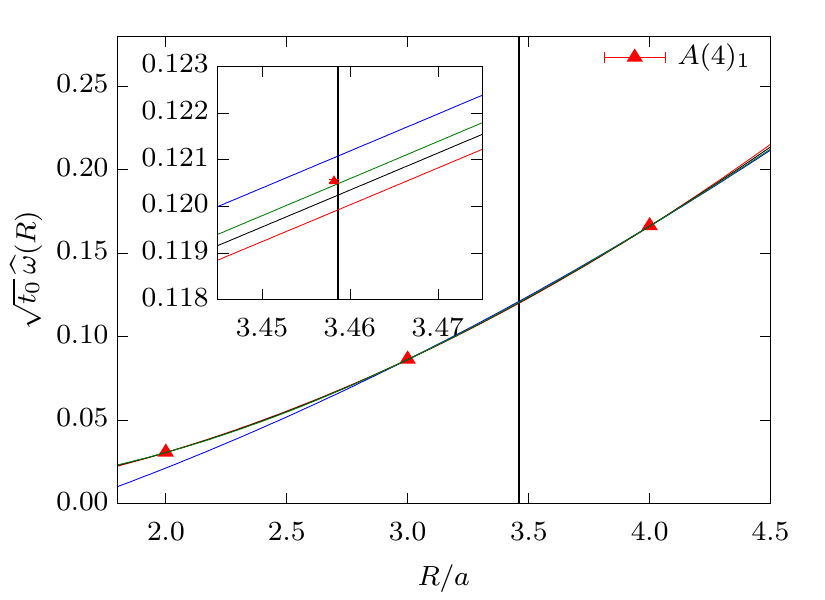}
	\includegraphics[width=0.49\textwidth]{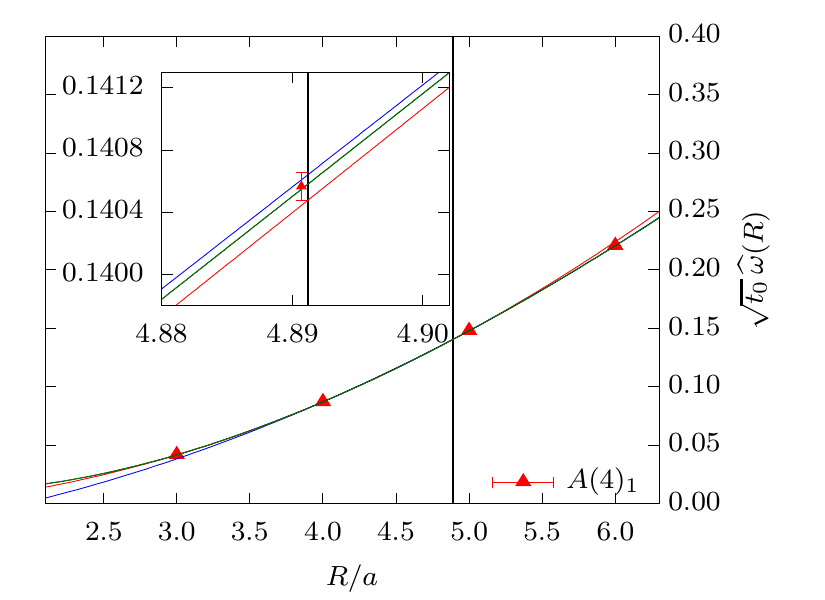}
	\caption{Systematic and statistical errors from the 
		interpolation in $R$ for the ensemble $A(4)_1$, which corresponds to 
		the gauge group \SU{4} at a lattice spacing $a \approx 0.1 \, \fm$. On the left, for $c=1/2$
, and on the right for $c=1$. In the first case, the statistical error
is barely visible in comparison to the systematic one, while in the 
second case, they are of comparable size.}
	\label{fig:interpsystem}
\end{figure}

Another potential source of systematic errors has to do with the interpolation in the 
flow time $t$ and in the loop size $R$. In the case of the flow time, the observables 
are measured in $t$ with a resolution of $0.02$ in units of $a^2$, so we find the effects of  
this interpolation to be negligible in comparison to the statistical errors. 
Concerning $R$, the situation is different and we must be careful 
with the systematics coming from this interpolation. As mentioned in the previous section,
in order to match the loops 
at different $a$ and different $\Nc$, we interpolate in their size $R$ 
to a value given by $R_c=\sqrt{8ct_0}$. To assess the systematic error 
from the interpolation we fit the data to a polynomial in the variable 
$\hat{\omega}(R) = -\frac{1}{R}\log{W(R)}$, where the dependence in the 
flow time $t$ has been omitted to simplify the notation. The fitting  
is done using two quadratic and two cubic functions of $R$ varying
the points used for the fit. The effects of the systematics from the 
interpolation are shown in \Fig\ref{fig:interpsystem}. We present two 
different cases; on the left, when $R_c/a \approx 3.5$, and on the right
, when $R_c/a \approx 4.9$. Clearly, when interpolating to a half 
integer value, the systematics from the interpolation are much larger 
than when interpolating to an almost integer value. Using the results 
from the fits, the central value is defined as 

\begin{equation}\label{eq:Fitval}
	W(c) = 1/2 \left( \max \left\lbrace W_1,W_2,W_3,W_4 \right\rbrace + 
	\min \left\lbrace W_1,W_2,W_3,W_4 \right\rbrace \right) \, ,
\end{equation}
where $W_i$ is the result from the $i-\text{th}$ fit. The systematic error $\Delta_S$ is defined as

\begin{equation}
\Delta_S=1/2 \left(\max \left\lbrace W_1,W_2,W_3,W_4 \right\rbrace - \min \left\lbrace W_1,W_2,W_3,W_4 \right\rbrace \right) \, .
\end{equation}
and it is combined in quadrature with the statistical one to obtain the 
final error at each point.

\section{Results}

\begin{figure}
   \centering
   \includegraphics[width=0.49\textwidth,clip]{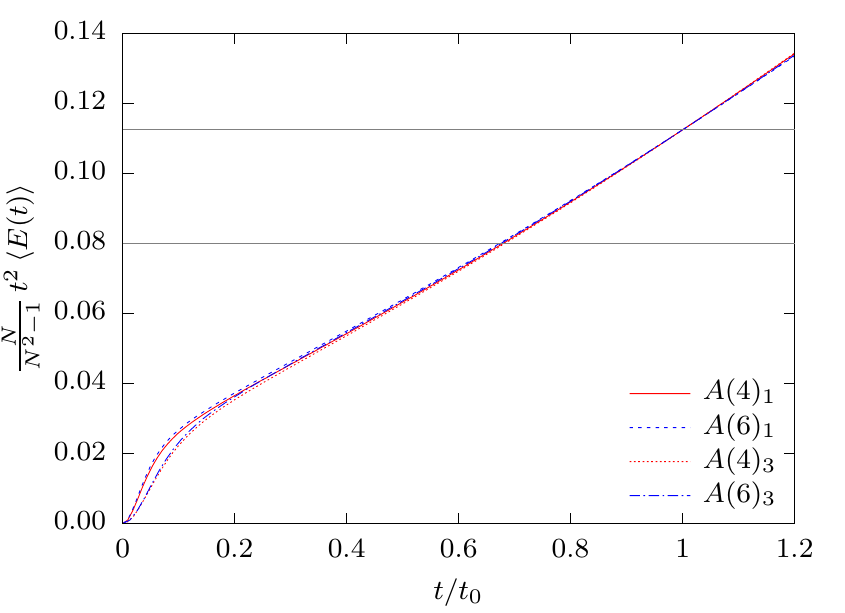}
   \includegraphics[width=0.49\textwidth,clip]{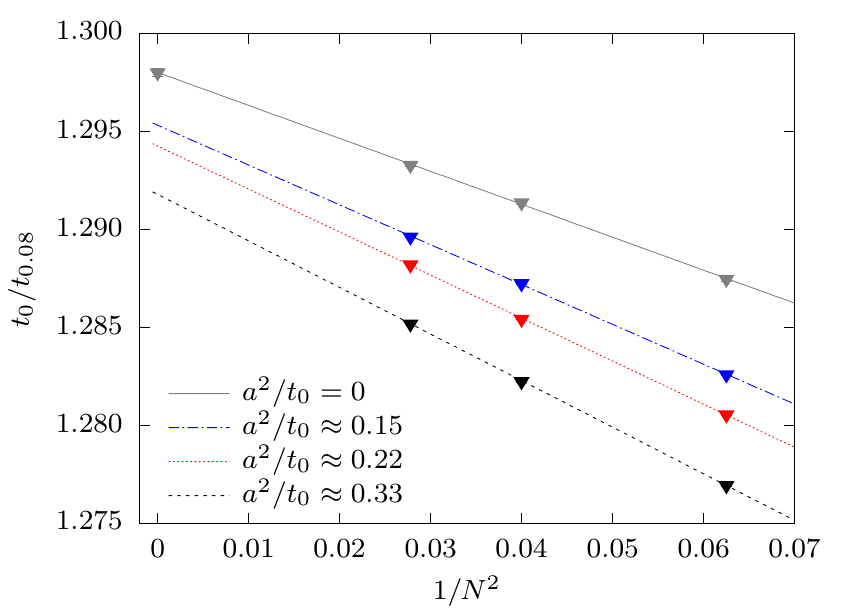}
   \caption{\textit{Left:} $t^2 \ev{E(t)}$ as a function of $t$ for several 
   gauge groups and lattice spacings. $A(4)_1$ corresponds to an \SU{4} 
   ensemble with $a \approx 0.1 \, \fm$, while $A(4)_3$ has a lattice spacing 
   $a \approx 0.067 \, \fm$. $A(6)_1$ and $A(6)_3$ correspond to \SU{6} 
   ensembles with the same lattice spacings as the \SU{4} ones correspondingly. 
   \textit{Right:} large $\Nc$ extrapolations of $t_0/t_{0.08}$ at three 
   different lattice spacings and in the continuum. Notice the excellent scaling 
   with $1/\Nc^2$ even at this very high accuracy.}
   \label{fig:t0vsN}
\end{figure}

Let us first discuss large $\Nc$ scaling. 
Before considering the loop observables, we use the scale 
itself. We can define 
a second scale similar to $t_0$ by changing the numerical pre-factor on the right 
hand side of \Eq\eqref{eq:t0def}. By replacing $0.1125$ with $0.08$ we 
have a perfectly reasonable scale, which we denote by $t_{0.08}$. As 
shown on the left panel of \Fig\ref{fig:t0vsN}, choosing the value of $0.08$ 
we are in the same region where $t^2 \ev{E(t)}$ grows roughly linearly with $t$ 
as is the case when $t$ is close to $t_0$. Notice that the $\Nc$ dependence is barely 
visible at the scale of the plot, while on the other hand cut-off effects are large in 
the region of small $t$.

On the right panel of \Fig\ref{fig:t0vsN} we show the large $\Nc$ extrapolations of 
$t_0/t_{0.08}$ and we observe an excellent agreement with a fit to a linear function in 
$1/\Nc^2$. For the three lattice spacings considered, we find the values of $\chidof$ 
to be equal to $0.35, \, 2.18$ and $1.30$ respectively. In the continuum, the 
fit is also excellent with a value of $\chidof = 0.96$ and a result for the large $\Nc$ 
and continuum extrapolation of $t_0/t_{0.08} = 1.29802(19)$. These results are in 
complete agreement with the \tHooft scaling. They test it with very high accuracies, 
as the errors in our measurements are $\rmO(10^{-4})$ (around $10000$ measurements). 
This allows us to verify the $1/\Nc^2$ scaling for this observable even when the 
finite $\Nc$ corrections are in the percent level.

\begin{figure}
   \centering
   \includegraphics[width=0.98\textwidth,clip]{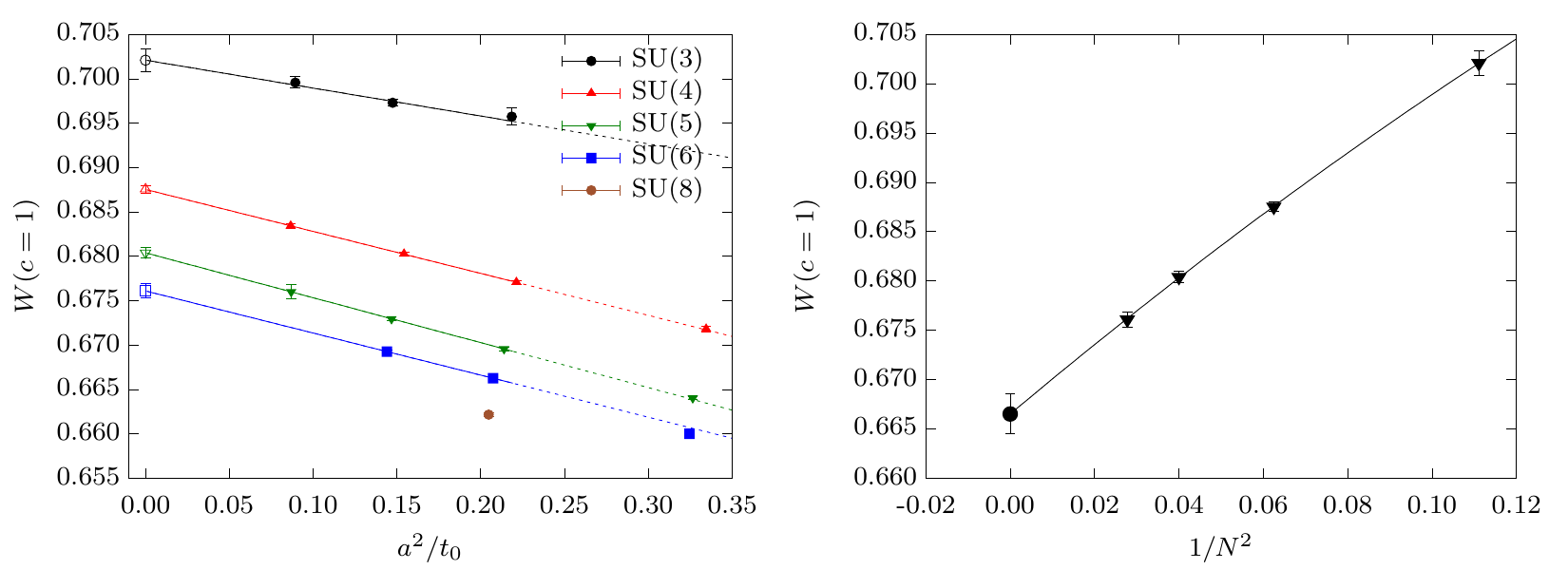}
   \caption{\textit{Left:} Continuum limit extrapolations of $W(c=1)$. 
   \textit{Right:} Large $\Nc$ extrapolation of $W(c=1)$ in the continuum. The results
   are perfectly compatible with a polynomial in $1/\Nc^2$ as expected from the 
   \tHooft topological expansion.}
   \label{fig:loopsvsN}
\end{figure}

Let us now turn to the smooth Wilson loop operators. In \Fig\ref{fig:loopsvsN} we show 
the continuum and large $\Nc$ extrapolations of $W(c=1)$. For the continuum 
extrapolations, we do a linear fit in $a^2/t_0$ using only the finer lattices, i.e., 
those for which $a^2/t_0 < 0.25$. With this choice, we take the continuum limit 
for \SU{3}, \SU{4} and \SU{5} using three data points, while we have only two points for 
\SU{6}. To assess the validity of this choice, we also perform a fit including the 
coarsest data 
points. We find the two strategies to give compatible extrapolations, so we decide 
to use the one with fewer points to trade an increase of the statistical error, with 
the reduction of the systematics due to neglecting higher order terms in the $a$-expansion. The fits are excellent as is displayed on the 
left panel 
of \Fig\ref{fig:loopsvsN}, and we find similar results for the cases $c=1/2$ and $c=9/4$. 
Notice that in the case of \SU{8} we have a single point, so we cannot take the 
continuum limit; hence the right panel in \Fig\ref{fig:loopsvsN} displays the large 
$\Nc$ extrapolation of the data in the continuum with
$\Nc \leq 6$. The large $\Nc$ fit is performed using a quadratic function in $1/\Nc^2$, 
with a value of $\chidof < 0.1$ and a small $1/N^4$ coefficient.

\subsection{Factorization}

\begin{figure}
   \centering
   \includegraphics[width=0.49\textwidth,clip]{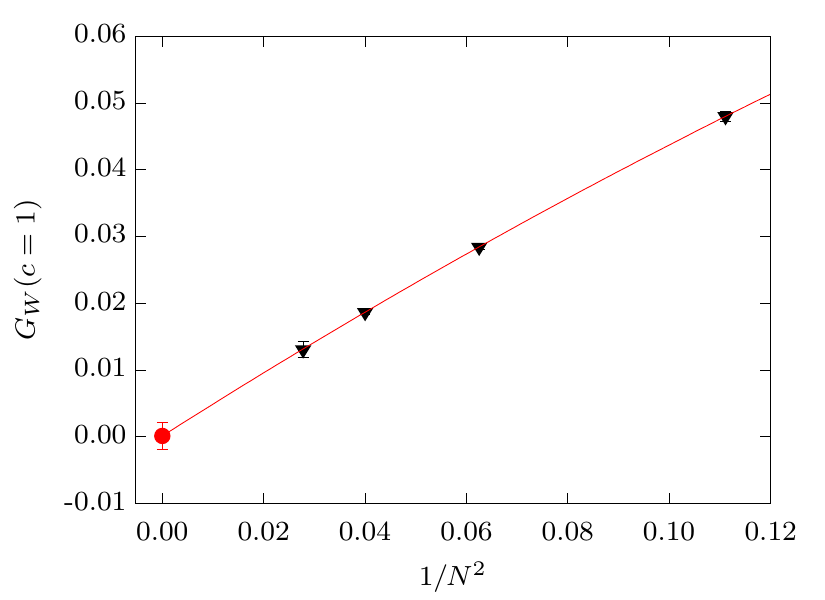}
   \includegraphics[width=0.49\textwidth,clip]{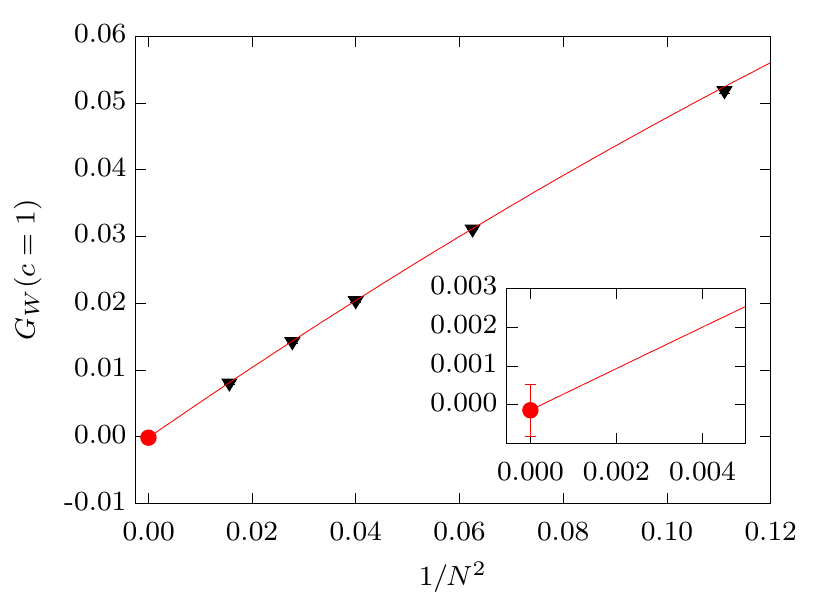}
   \caption{Large $\Nc$ extrapolation of $G_W(c=1)$. \textit{Left:} In the continuum.
   \textit{Right:} At finite lattice spacing. In both cases, the fits are done 
   without including the \SU{3} point.}
   \label{fig:GWvsN}
\end{figure}

In order to verify the property of factorization, \Eq\eqref{eq:factorization},
we check whether $G_W$, \Eq\eqref{eq:Glattice}, satisfies $G_W \to 0$ when $1/\Nc \to 0$. We take the continuum limits for all $\Nc$ except for 
$\Nc=8$. As before, those lattices for which 
$a^2/t_0 > 0.25$ are used only for validation. We also interpolate 
to fixed $a^2 /t_0$ given by the one of
the \SU{8} ensemble .  We can then also
use the \SU{8} point for the large $\Nc$ extrapolations at a fixed $a$ ($a^2 / t_0 \approx 0.21$).
In addition to having an extra point at the larger value of $\Nc=8$, by 
working at finite lattice spacing, only an interpolation is required for 
$\Nc<8$, yielding  smaller errors. Of course one must keep in mind that the finite $a$ results
are not universal; they depend on the regularization, here the Wilson plaquette
action. Still, they can be used as a test of large $\Nc$ scaling.
Graphs of the large $\Nc$ 
extrapolations at finite lattice spacing and in the continuum are presented in 
\Fig\ref{fig:GWvsN}. 

We observe that a quadratic fit excluding \SU{3} correctly 
extrapolates to $1/\Nc = 1/3$ within two, very small, standard deviations at all $c$. 
While a priori it can't be expected that the expansion works so well
also for $\Nc=3$, this fact serves to increase our confidence 
in the large $\Nc$ fit. The results of the quadratic fit excluding \SU{3} 
are presented in \Tab\ref{tab:largeNres}. As shown, the results of 
the large $\Nc$ extrapolation agree with zero and thus support factorization at all 
values of $c$. 
To further validate this conclusion, we perform a fit to the data forcing it 
to pass through zero when $1/\Nc =0$. The value of $\chidof$ of the fits can then be used as a criterion to test factorization. For different values of $c$ we present the 
results for $\chidof$ at finite lattice spacing of these constrained fits in 
\Tab\ref{tab:largeNres}. All of them are excellent and support
factorization.

\begin{table}
	\centering
	\caption{Results of the large $\Nc$ extrapolations, $~^{(1)}$ in the 
	continuum, $~^{(2)}$ at finite lattice spacing; and $~^{(3)}$, at finite 
	lattice spacing with the constrained fit.}
	\label{tab:largeNres}
	\begin{tabular}{cccc}
	           	& $c=1/2$ & $c=1$ & $c=9/4$ \\\toprule
$G_W(1/\Nc =0)^{(1)}$ 	& $-0.0001(54)$ & $-0.0004(73)$ & $0.010(13)$\\\midrule
$G_W(1/\Nc =0)^{(2)}$   & $0.00001(63)$ & $-0.0001(6)$  & $-0.00074(96)$ \\
$\chidof^{(2)}$	        & $<0.01$ & $0.01$ & $0.18$ \\\midrule
$\chidof^{(3)}$	        & $<0.01$ & $0.03$ & $0.38$ \\\bottomrule
	\end{tabular}
\end{table}

\begin{figure}
   \centering
   \includegraphics[width=0.49\textwidth,clip]{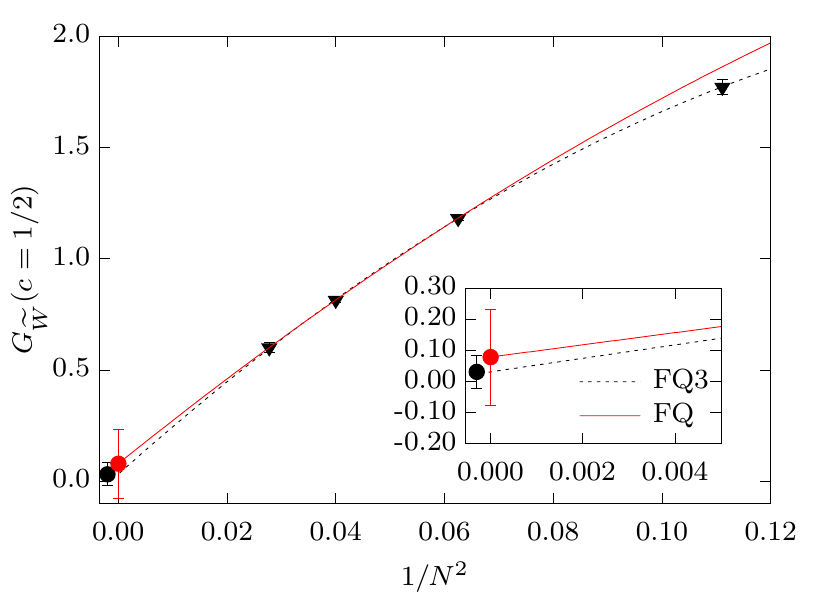}
   \includegraphics[width=0.49\textwidth,clip]{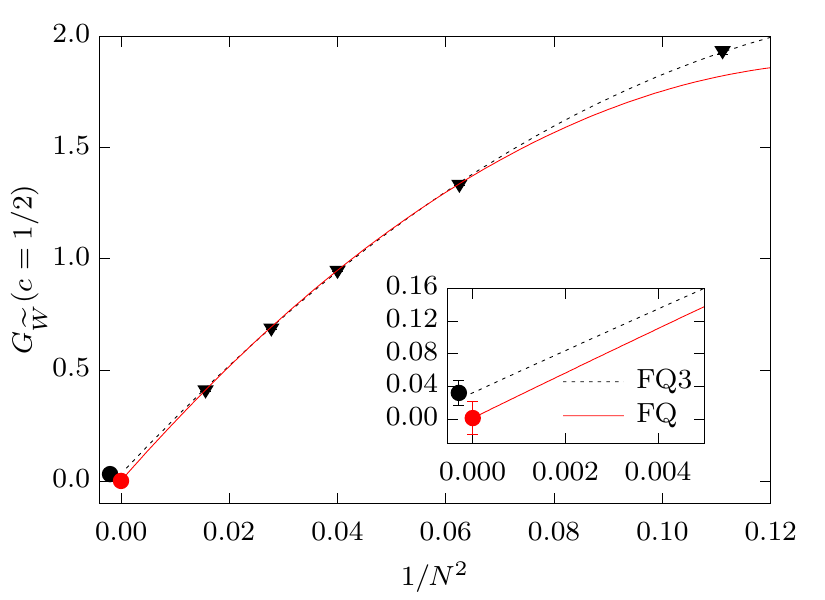}
   \caption{$G_{\widetilde{W}}$ as a function of $\Nc$. On the left in the continuum 
   and on the right at finite lattice spacing. In both cases, we show the fit excluding 
   the \SU{3} point (FQ), and the one including it (FQ3). It is clear that the 
   $\rmO(1/\Nc^4)$ corrections are required to include the \SU{3} point and it 
   has to be included in the fit.  Both at finite 
   lattice spacing as in the continuum, the large $\Nc$ extrapolation is compatible 
   with factorization.}
   \label{fig:GWRc2}
\end{figure}

So far we have only considered the case $R_c = \sqrt{8ct_0}$. To study what 
happens at different values of $R$ we decided to use the loops measured at $c=1/2$ 
and fix their size such that $\tilde{R}_c = 2 \sqrt{8ct_0}$. Let us denote this observable 
as $G_{\widetilde{W}}$. The analysis is then the same 
as described earlier except that in this case, when excluding \SU{3} from the fits, 
the resulting function does not extrapolate to \SU{3}. This, however, does not modify the main 
conclusion, and taking the large $\Nc$ limit shows that factorization also holds. 
A plot showing our results in the continuum and at finite lattice spacing is displayed in 
\Fig\ref{fig:GWRc2}. 

Finally, we also considered a more complicated observable constructed from 
the smooth Wilson loops. For that, we define

\begin{equation}
	W^2_s (c) = \frac{a}{(T-2d)} \sum_{x_0 = d}^{T-d-a} 
	\left( \frac{a^3}{L^3}  
	\sum_{\vec{x}} W(ct_0,x_0,\vec{x},R_c) \right)^2 \, ,
\end{equation}
so that 

\begin{equation}
	H_W(c) = \left( \frac{L^3}{t_0^{3/2}} \right) 
	\frac{\ev{W^2_s(c)} - \ev{W(c)}^2}{\ev{W(c)}^2} \,  ,
	\label{eq:HWlattice}
\end{equation}

\begin{figure}
   \centering
   \includegraphics[width=0.49\textwidth,clip]{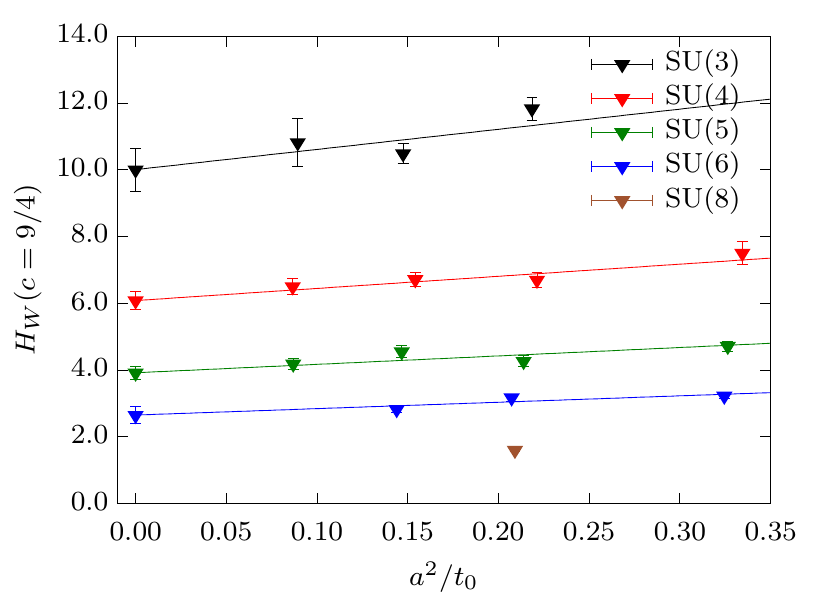}
   \includegraphics[width=0.49\textwidth,clip]{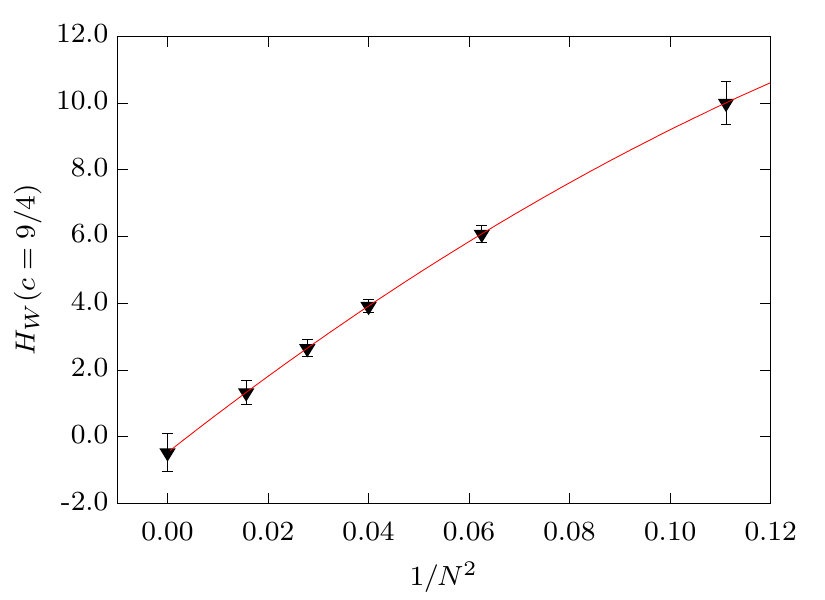}
   \caption{Continuum (left) and large $\Nc$ (right) extrapolations of 
   $H_W$ performing a global fit to the data.}
   \label{fig:HWvsN}
\end{figure}

Once again, $H_W(c)$ can be used to test factorization, but due 
to significant finite volume effects for $c\leq1$, we can only estimate the errors 
reliably for $c=9/4$. Once more, the results for 
this observable are compatible with factorization. In this case, we 
performed a global fit to the data ($\chidof = 1.73$) including the 
corrections of $\rmO(1/\Nc^2)$, 
$\rmO(a^2)$, $\rmO(a^2/\Nc^2)$ and $\rmO(1/\Nc^4)$. The results are shown in 
\Fig\ref{fig:HWvsN}.

%


\section{Conclusions}

We have used high precision data to test 
the scaling in powers of $1/\Nc^2$ predicted by the \tHooft expansion 
for the pure gauge theory. By using the gradient flow, we were able to use 
the smooth Wilson loops and study their $\Nc$ dependence both at finite lattice 
spacing and in the continuum. In both cases, we have found that the observables 
are extrapolated to $1/\Nc = 0$ in powers of $1/\Nc^2$ as expected. 
Moreover, by using high precision data, we can observe the $\Nc$ 
dependence at below the percent level. For the observables that we have considered, 
we find that corrections of $\rmO(1/\Nc^2)$ describe very well the data for 
$\Nc >3$, while including the results at $\Nc =3$ generally requires the addition 
of a term of $\rmO(1/\Nc^4)$.

We have further presented, to our knowledge, the first direct non-perturbative verification 
of large $\Nc$ factorization for \SU{\Nc} gauge theories on the lattice. 
While factorization, \Eq\eqref{eq:factorization}, holds within our small 
uncertainties, the corrections at finite $N=3-10$ can be very large, see figs~\ref{fig:GWvsN},\ref{fig:GWRc2}.
Note that $\tilde G=1$ or $H=1$ means there is a 100\% violation of factorization.
These large $1/N^2$ corrections are very likely related to the very large $N$ values needed to approach the 
$1/N=0$ limit in the one-point model \cite{GonzalezArroyo:2012fx}.
In particular, the increase of the $1/N^2$ corrections with the 
size of the loops, which we observe in Figs. \ref{fig:GWvsN} and \ref{fig:GWRc2}, 
is present in the one-point models~\cite{Bringoltz:2011by}.

\vspace{1cm}
\begin{acknowledgement}
\textbf{Acknowledgements.} We would like to thank M. C\`e, L. Giusti and 
S. Schaefer 
for sharing part of the generation of the gauge configurations \cite{Ce:2016awn}
and M. Koren for useful discussions. 
Our simulations were performed at the ZIB computer center with the 
computer resources granted by The North-German 
Supercomputing Alliance (HLRN). 
M.G.V acknowledges the 
support from the Research Training Group GRK1504/2 ``Mass, Spectrum, Symmetry'' 
funded by the German Research Foundation (DFG). 
\end{acknowledgement}

\providecommand{\href}[2]{#2}\begingroup\raggedright\endgroup


\end{document}